\documentclass{sf2a-conf2025}
\usepackage{graphicx}
\usepackage{hyperref}
\usepackage[]{natbib}  
\usepackage{epstopdf}

\def\BibTeX{{\rm B\kern-.05em{\sc i\kern-.025em b}\kern-.08em
    T\kern-.1667em\lower.7ex\hbox{E}\kern-.125emX}}
\bibpunct{(}{)}{;}{a}{}{,}  

\newcommand{\tcoe}{\mbox{tCO$_2$e}}
\newcommand{\Mtcoe}{\mbox{MtCO$_2$e}}
\newcommand{\Mtcoeyr}{\mbox{MtCO$_2$e yr$^{-1}$}}

\newcommand{\knodlseder}{\mbox{J. Kn$\ddot{\mbox{o}}$dlseder}}

\begin{document}

\TitreGlobal{SF2A 2025}
\title{Environmental impacts of astronomical research infrastructures}
\runningtitle{Environmental impacts of astronomical research infrastructures}

\author{\knodlseder}\address{Institut de Recherche en Astrophysique et Plan\'etologie, Universit\'e de Toulouse, CNRS, CNES, UPS,
9 avenue Colonel Roche, 31028 Toulouse, Cedex 4, France}

\setcounter{page}{237}

\maketitle

\begin{abstract}
Human activities degrade the Earth environment at an unprecedented scale and pace, threatening 
Earth-system stability, resilience and life-support functions.
We can of course deny the facts, get angry about them, or try to bargain.
Or we may overcome these stages of grief and move towards accepting that human activities need to 
change, including our own ones.
The purpose of this paper is to support astronomers in this transition, by providing insights into the 
origins of environmental impacts in astronomical research and proposing changes that would make 
the field sustainable.
The paper focuses on the environmental impacts of research infrastructures, since these are
the dominant sources of greenhouse gas emissions in astronomy, acknowledging that impact
reductions in other areas, for example professional air travelling, need also to be achieved.
\end{abstract}



\section{Introduction}

Carbon footprint estimates suggest that research infrastructures are the largest contributor to the
greenhouse gas emissions of astronomical research \citep{martin2022,knodlseder2022}.
What is called here research infrastructures are the astronomical observatories built on Earth and 
the telescopes and probes that are sent to space and that we use to learn about the Univers.
Recent research suggests that the annual greenhouse gas emissions of astronomical facilities reached 
a value of 1.3 \Mtcoeyr\ in 2022, of which 84\% stem from space missions and 16\% from 
ground-based observatories \citep{knodlseder2024}.
The unit of \Mtcoe\ refers here to emissions that are equivalent to one million metric tons of CO$_2$,
which is the usual unit that is used to aggregate different greenhouse gases into a single number.
The research also revealed that the number of operating facilities has been constantly growing, with 
an average growth rate of 3.2\% per year for space facilities and 1\% per year for ground-based 
facilities over the past 30 years \citep{knodlseder2024}.
The question hence arises of whether these growth rates are environmentally sustainable.

Much of the recent research has focused on quantifying greenhouse gas emissions of 
astronomical research that contribute to climate change, yet human activities have additional
detrimental environmental impacts that need also to be assessed and reduced.
Earth system science has identified nine processes and associated boundaries that, if crossed, 
could generate unacceptable environmental change that threatens Earth-system stability, resilience
and life-support functions.
Six of these nine planetary boundaries have already been breached, including 
climate change, 
rate of biodiversity loss, 
interference with the nitrogen and phosphorus cycles, 
global freshwater use, 
change in land use, and
chemical pollutions
\citep{caesar2024}.
Recent research suggests that with ocean acidification a seventh planetary boundary has already
been crossed, leaving stratospheric ozone depletion and atmospheric aerosol loading as the only
processes that are still within the safe operating space \citep{findlay2025}.

In order to gain insights into all environmental impacts of astronomical facilities we have conducted 
Life Cycle Assessments (LCAs) for a space system and a ground-based
facility \citep{barret2024,dosSantosIlha2024}.
Furthermore, we have investigated how greenhouse gas emissions from astronomical facilities
may evolve in the future, depending on the choices we make as a community and of funding
agencies \citep{knodlseder2024}.
In this paper we summarise the results of these studies and discuss their implications on the
sustainability of our research activities.

\section{Estimating and reducing environmental impacts using Life Cycle Assessment and eco-design}

\subsection{Development and construction of X-IFU instrument}

\begin{figure}[ht!]
\centering
\includegraphics[width=0.8\textwidth,clip]{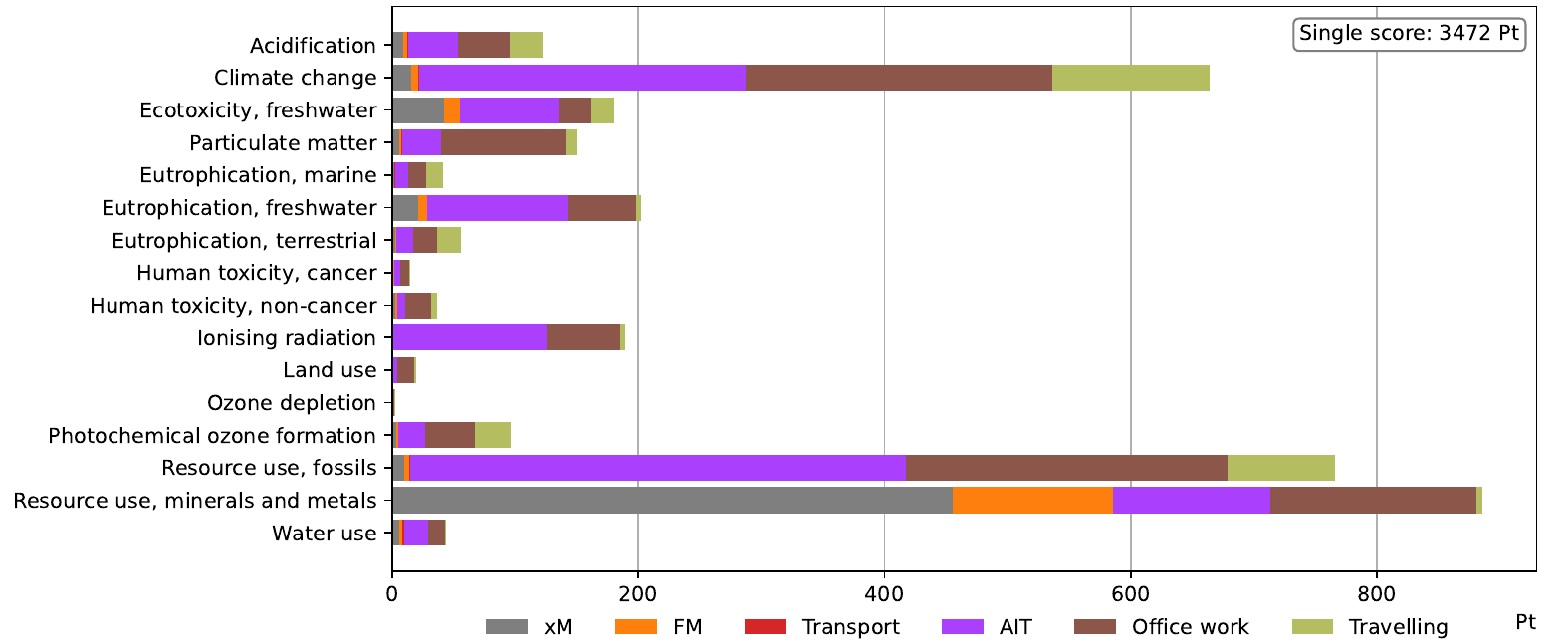}    
\caption{
Life cycle assessment results for the development and construction of the X-IFU instrument aboard
ESA's Athena mission \citep{barret2024}.
Impact categories are shown on the ordinate, impacts are given on the abscissa and are expressed
in units of ``points'' (Pt), where one point corresponds to the annual environmental impact of one human.
Impacts for the different impact categories can be added together.
The sum of all impacts results in a ``single score'' that expresses the total environmental impact of
developing and constructing X-IFU.
The different colours show different activities, comprising
development and construction of pre-flight models (xM),
development and construction of flight models (FM),
transport of system elements,
assembly, integration and testing (AIT),
office work and travelling.
}
\label{fig:xifu-lca}
\end{figure}

The first LCA we made concerns the X-ray Integral Field Unit (X-IFU), a cryogenic instrument that will 
provide from 2035 on high-resolution X-ray spectroscopy aboard the Athena space observatory of 
the European Space Agency (ESA) \citep{barret2024}.
The weight of the considered instrument is 221 kg, LCA results are summarised in Fig.~\ref{fig:xifu-lca}.
The most significant environmental impacts are ``Climate change''  and ``Resource use'', including
fossil, mineral and metal resources.
Total greenhouse gas emissions amount to $25\,536$ \tcoe\ with important contributions from
assembly, integration and testing (driven by energy consumption of clean rooms),
office work (driven by energy consumption of office buildings and staff commuting), and
travelling (driven by air travels for project meetings).
As most energy is produced from fossil resources their use shows a similar pattern.
The most important contribution to the use of mineral and metal resources comes from the construction 
of pre-flight and flight hardware, driven by the use of gold in the integrated circuits of the electronics.
The contribution from transport of system elements is negligible.

\subsection{Construction and operations of CTAO mid-sized telescope}

\begin{figure}[ht!]
\centering
\includegraphics[width=0.48\textwidth,clip]{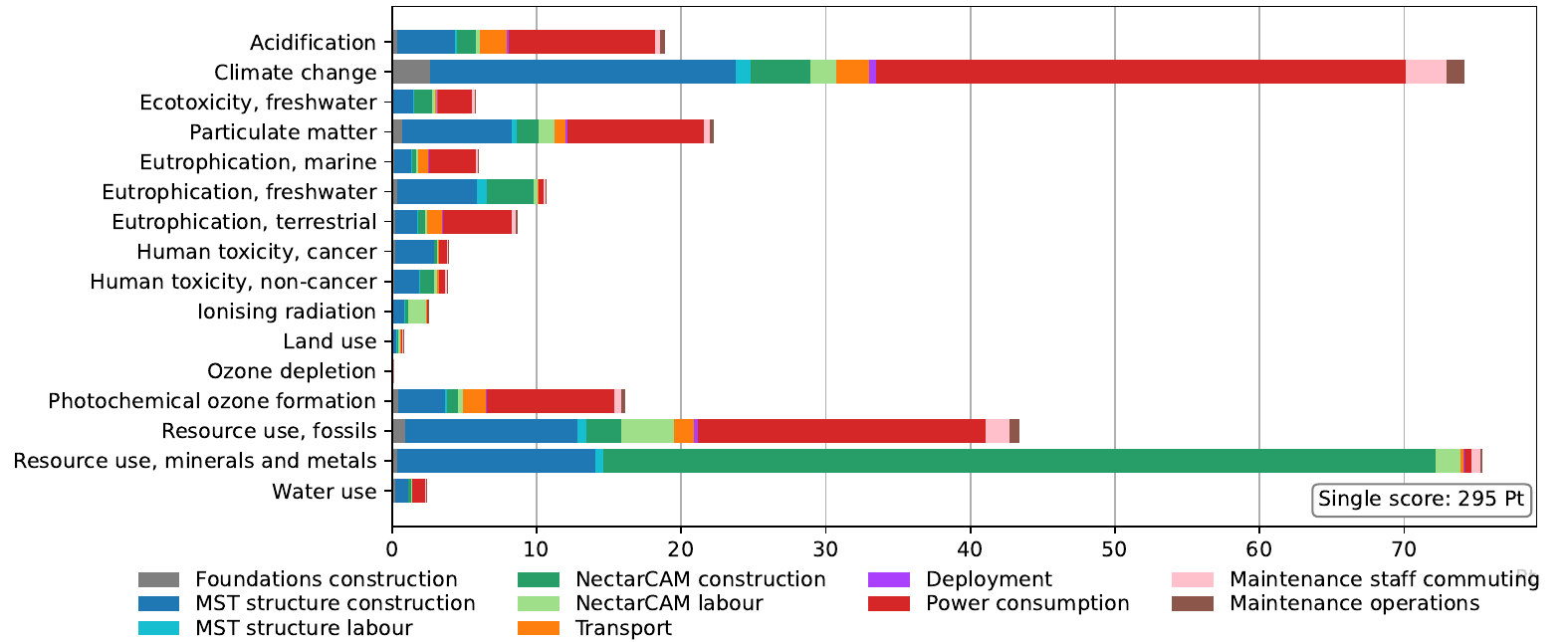}    
\includegraphics[width=0.48\textwidth,clip]{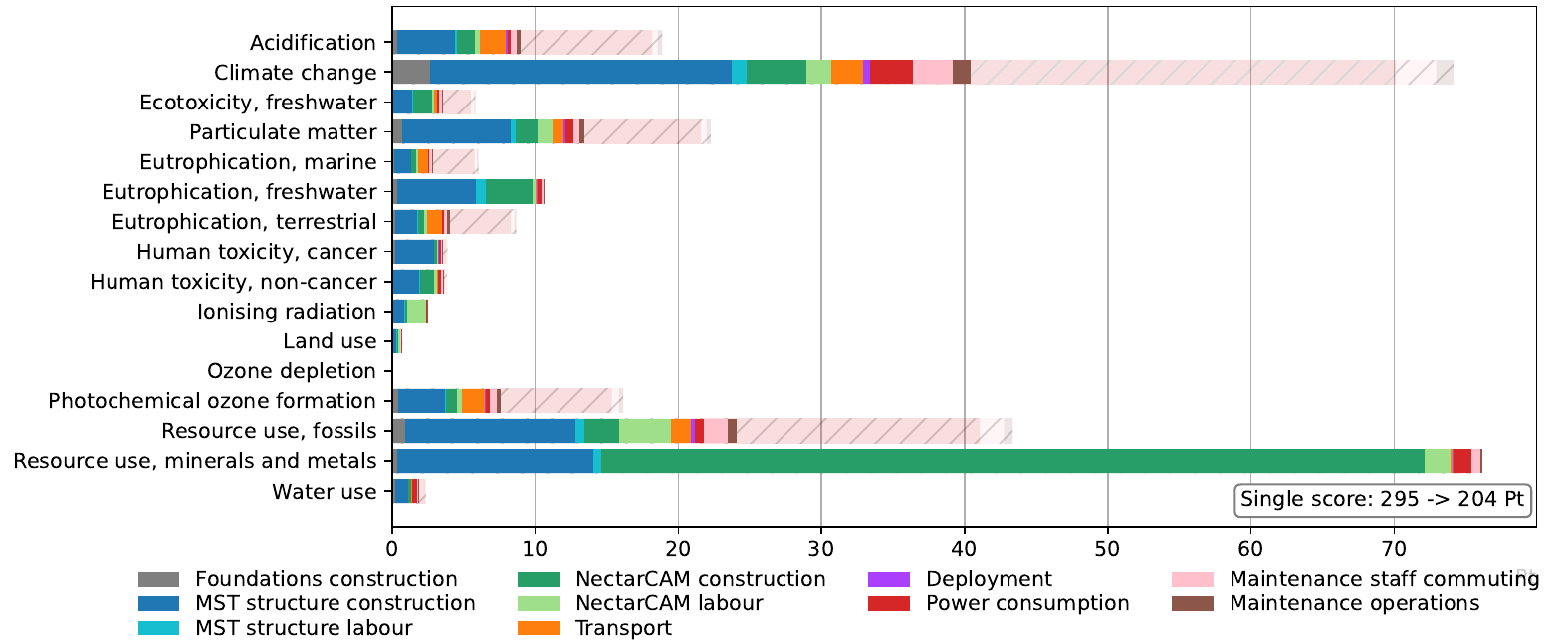}    
\caption{
{\bf Left:} Life cycle assessment results for the construction and operations of the CTAO mid-sized 
telescope on the site of La Palma \citep{dosSantosIlha2024}.
The different colours show different activities, comprising
construction of
foundations, telescope structure (MST) and camera (NectarCAM),
related labour,
transport of systems,
deployment of the telescope on site,
power consumption during operations,
and maintenance commuting and operations.
{\bf Right:} Results for an alternative model where diesel generators were replaced by a wind turbine 
powered hydro storage system for electricity production on La Palma. 
Impact reductions are indicated by the hatched light portion of the horizontal bars.
}
\label{fig:mstn-lca}
\end{figure}

The second LCA we made concerns the construction, upgrade and 30 years of operations of one Mid-Sized 
Telescope (MST) of the Cherenkov Telescope Array Observatory (CTAO) that will be installed in
the coming years on the island of La Palma, Spain \citep{dosSantosIlha2024}.
The telescope consists of foundations (464 t), a structure including the mirrors (82 t), and the 
NectarCAM camera (2.1 t; all weights are in metric tons).
LCA results are summarised in the left panel of Fig.~\ref{fig:mstn-lca}.
The most significant environmental impacts are again ``Climate change''  and use 
of mineral and metal resources.
Total greenhouse gas emissions amount to $2\,660\pm274$ \tcoe\ with important contributions from the 
construction of the telescope structure (driven by the production and processing of steel and 
aluminium) and the powering of the telescope during the 30 years of operations (driven by
electricity generation on La Palma using diesel generators).
The most important contribution to the use of mineral and metal resources comes from
camera construction (driven by the use of gold in integrated circuits), and to a lesser extent
from the construction of the telescope structure.
Construction of the telescope foundations, transport of the telescope on site, telescope deployment 
and maintenance contribute little to the environmental impacts.

\subsection{Reducing environmental impacts using eco-design}

Using the LCA model of the CTAO mid-sized telescope we investigated whether environmental
impacts can be reduced by making changes to the design or operations of the telescope.
Replacing for example the steel-made counterweights by concrete blocks of the same weight would 
reduce the environmental impact (as measured by the single score) by 3\%, with main reductions 
achieved on climate change and fossil resource use.
Using metal machining instead of metal casting for the production of the head and yokes of the 
telescope positioner would reduce the single score by 12\%, with impact reductions between 5\% 
and 25\%, depending on impact category.
Specifically, the impact on climate change would be reduced by 14\%, corresponding to greenhouse
gas emission reductions of 381 \tcoe.
The most dramatic impact reductions would be achieved if the diesel generators for electricity 
generation on La Palma were replaced by a wind turbine powered hydro storage system alike the 
one that is currently operating on the neighbouring island of El Hierro  
(see right panel of Fig.~\ref{fig:mstn-lca}).
While the construction of such a system would increase the environmental impact of mineral resource 
use by 1\%, all other impacts would be reduced, with impact reductions of up to 54\% on some 
categories.
The single score would be reduced by 31\%, and greenhouse gas emissions would be lowered by 
46\%, corresponding to $1\,211$ \tcoe.

\section{Choosing the greenhouse gas emission pathway}

\begin{figure}[ht!]
\centering
\includegraphics[width=0.74\textwidth,clip]{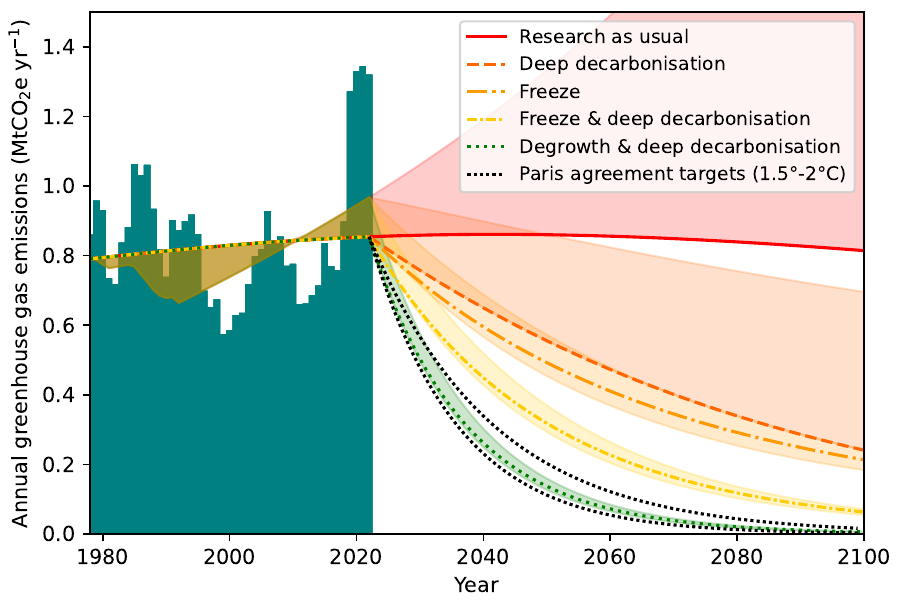}    
\caption{
Evolution of annual greenhouse gas emissions from astronomical research infrastructures under
different policy choices, including freezing or decreasing the number of facilities and increasing their
decarbonisation \citep[adapted from][]{knodlseder2024}.
Evolutions were based on current trends, with lines obtained from trends over the last 45 years and 
shaded bands for shorter periods down to 30 years.
The histogram shows estimates of past greenhouse gas emissions for the period 1978--2022, where 
variations are due to construction activities of large space missions.
The recent peak in emissions is due to a surge of missions to the Moon.
The black-dotted lines show annual emission reductions of 5--7\%, which are the levels required
to meet the Paris agreement targets.
}
\label{fig:pathways}
\end{figure}

We studied the past evolution of the annual greenhouse gas emissions from astronomical research 
infrastructures based on an inventory of 586 ground-based facilities and 625 space missions, and
investigated how these emissions may evolve in the future based on possible policy choices
\citep[see Fig.~\ref{fig:pathways} and][]{knodlseder2024}.
Our work shows that if we choose to continue doing research as usual, emissions may either surge, 
or in the best case remain roughly constant, owing to an assumed (and so far observed) industrial 
decarbonisation rate of 1.4\% per year.
We could however alternatively choose to increase the decarbonisation rate, for example by asking
funding agencies to make substantial investments in the decarbonisation of existing facilities.
Assuming that ``deep decarbonisation'' levels of 3\% per year can be achieved, annual greenhouse 
gas emissions would drop, yet not at the pace required to meet the Paris agreement targets.
Alternatively we could choose to ``freeze'' the number of operating facilities to the current number,
which would give about the same result as the ``deep decarbonisation''.
Combining both policy options, which probably happens naturally if we choose to divert funding
from construction of new facilities to decarbonisation of existing facilities, would even more reduce 
the annual greenhouse gas emissions, yet still not at a rate that will halt global warming.
Only if we choose to degrow the number of operating facilities by 3\% per year and, at the same 
time, choose to invest in ``deep decarbonisation'', astronomical research would be on track towards 
sustainability.

\section{Towards sustainable astronomy}

Our research suggests that eco-design of astronomical facilities may reduce their environmental 
impacts, yet these reductions will probably not exceed a few $10\%$ at best.
Given the current annual growth rate of 1--3\% in the number of facilities, environmental impact 
reductions from eco-design will be quickly outpaced by the growth of the field.
As we demonstrated, without deliberately limiting the number of facilities and at the same time
investing in the decarbonisation of the field, astronomy will not become sustainable.
This does not imply that astronomical research will have to stop, nor that no new facilities could be 
built, but it implies that a more frugal approach needs to be developed.

How could astronomical research look like in the future?
There would be less competition among nations, regions, and scientists and more collaboration to 
make astronomy a common good instead of being a reaping contest.
This would imply less funding needs for new facilities and free resources for the decarbonisation of 
existing facilities, which would become the priority. 
Research would be mainly made using observations from existing facilities and archival data, and
slow science would be valued instead of citation counts.
Environmental impact budgets would be defined that diminish with time, making sure that astronomy 
stays on track on its way towards sustainability.
New facilities would only be built if existing datasets in related scientific fields have been fully exploited
and if their realisation does not breach the environmental impact budgets.
Through these changes we would illustrate how humanity can make peace with nature, making
astronomy once again a cultural good and no longer a showpiece of technological power.

\bibliographystyle{aa}  
\bibliography{knodlseder_S13} 

\end{document}